%% file: main.tex
\documentclass[authorversion,manuscript,screen,sigconf]{acmart}

\setcopyright{rightsretained}
\copyrightyear{2022}
\acmYear{2022}
\acmDOI{10.48550/arXiv.2208.11774}

\acmConference[SIGGRAPH Frontiers 2022]{SIGGRAPH 2022: ACM Computer Graphics and Interactive Techniques}{August, 2022}{Vancouver, Canada}
\acmBooktitle{SIGGRAPH Frontiers 2022: ACM Computer Graphics and Interactive Techniques, Frontiers workshop, 
August, 2022, Vancouver, Canada}
\usepackage[ruled, vlined]{algorithm2e}
\usepackage{wrapfig}
\begin{document}

\input{title-abstract.tex}


% \ccsdesc[500]{Human-centered computing~User models}

% \keywords{datasets, neural networks, gaze detection, text tagging}

%\input{teaser.tex}

\maketitle

\input{body.tex}

\bibliographystyle{ACM-Reference-Format}
\bibliography{main}

\end{document}

%% file: title-abstract.tex
% \title{Beyond Real-time Graphics: Rendering for Esports and High Refresh Rates}
\title{The Esports Frontier: Rendering for Competitive Games}

% 4 authors on one row
\settopmatter{authorsperrow=4}

\author{Josef Spjut}
\orcid{0000-0001-5483-7867}
% \email{jspjut@nvidia.com}
\affiliation{%
  \institution{NVIDIA}
  \country{USA}
}
\author{Arjun Madhusudan}
% \email{amadhusudan@ncsu.edu}
\affiliation{%
  \institution{NCSU \and NVIDIA}
  \country{USA}
}
\author{Benjamin Watson}
\orcid{0000-0002-3758-7357}
% \email{bwatson@ncsu.edu}
\affiliation{%
  \institution{NCSU}
  \country{USA}
}
\author{Ben Boudaoud}
\orcid{0000-0003-4195-0793}
% \email{bboudaoud@nvidia.com}
\author{Joohwan Kim}
% \email{sckim@nvidia.com}

\affiliation{%
  \institution{NVIDIA}
  \country{USA}
}
\renewcommand{\shortauthors}{Spjut et al.}

\renewcommand{\shorttitle}{The Esports Frontier}

\begin{abstract}
Real-time graphics is commonly thought of as anything exceeding about 30 fps, where the interactivity of the application becomes fluid enough for high rates of interaction.
Inspired by esports and competitive gaming, where players regularly exceed the threshold for real-time by 10x (esports displays commonly reach 360 Hz or beyond), this talk begins the exploration of how rendering has the opportunity to evolve beyond the current state of focus on either image quality or frame rate.
Esports gamers regularly decline nearly all options for increased image quality in exchange for maximum frame rates.
However, there remains a distinct opportunity to move beyond the focus on video as a sequence of images and instead rethink the pipeline for more continuous updates.
\end{abstract}

%% file: body.tex
\section{Introduction}

Esports gamers play to win and gain much of their enjoyment from prevailing over intense competition.
To them, every millisecond matters and they are willing to sacrifice many conveniences and graphics features if it will help them win.
There are copious examples of these types of gamers turning down the deliberately designed graphical features that were so carefully and thoughtfully put into these games, sometimes for minimal performance improvement.
Graphics researchers could look at esports gamers and decide that since they are not very interested in photo-realistic rendering, they are not worth serving.
Gamers would certainly benefit by better understanding which features actually impact their performance and which are merely placebo.
We argue instead that esports gaming is a distinct opportunity for graphics researchers to create new rendering systems, intentionally designed for this performance-minded demographic.

In particular, we focus on three areas of opportunity that are under-explored by the computer graphics community.

\begin{enumerate}
    \item High frame rate rendering
    \item Low latency rendering
    \item Rendering for maximum perceptual performance
\end{enumerate}

While high frame rate and low latency often go hand-in-hand, there are distinct reasons why each may benefit the player.
Additionally, certain rendering architecture approaches may actually add latency in exchange for increased frame rate, and it is essential to consider these trade offs.

\section{Background}

While not all interactive applications depend on responsiveness and maximizing performance, competitive video games are a primary example of these increasingly important systems.
Esports games are the most popular and long-lived games on the market, and consistently rank among the most played games year after year.
Like ball games that acquire history and tradition, supporting development of strategy and tactics over the years with consistent rule sets; so esports also change slowly.
Traditional sports players can choose some equipment, such as shoes and clothing. Similarly, esports players can choose computer hardware and configure game settings.

For example, many game developers use an approach to synchronize game frame completion with the scan-out of information to the display, often called VSync.
When VSync is off, the frames displayed on the screen often contain a visual artifact known as screen tearing, when the top and bottom of the image come from two separately rendered frames of the game.
% TODO: Arjun?
In a recent survey of Dota 2 players, Madhusudan and Watson~\cite{madhusudan2021better} found that over 80\% of players turn off VSYNC.
This player preference may be a result of a side-effect of turning VSync on, which is that the end-to-end latency of the game can be a full frame longer than when VSync is off (e.g. 16.67 ms at 60 Hz).
% VSYNC is commonly forced on in most console games. 
% It prevents screen tearing by synchronizing the swap of frames with the display scan-out, and is a basic step toward high fidelity graphics.
% VSYNC's synchronization requires the game to wait for the display scan-out to begin before swapping in the new frame, and so introduces latency and reduces the responsiveness to user inputs.
Madhusudan and Watson also found that a majority of users turned off several other graphics features to improve computer performance and reduce latency, including shadows, texture detail, and particle effects.

% Talk about user's preferences to turn down graphics. \cite{madhusudan2021better}

\section{High frame rate rendering}

\begin{figure}
    \centering
    \includegraphics[width=0.31\columnwidth]{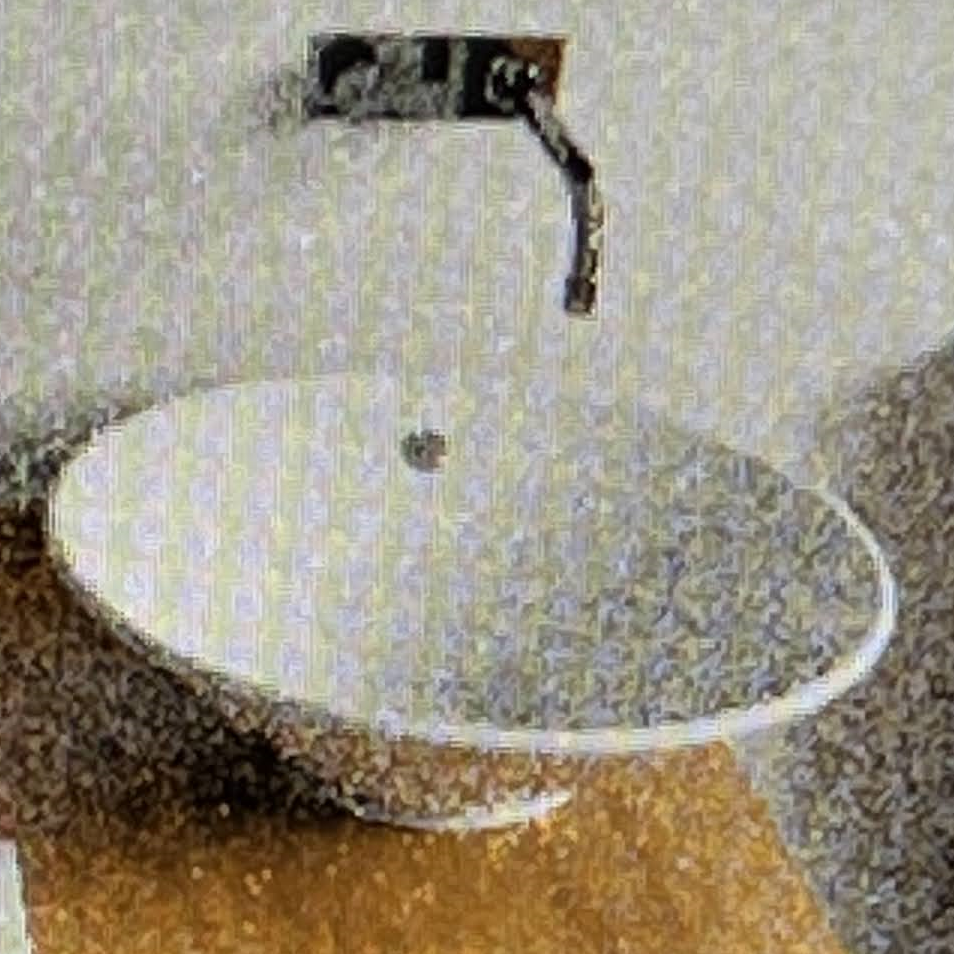}
    \includegraphics[width=0.31\columnwidth]{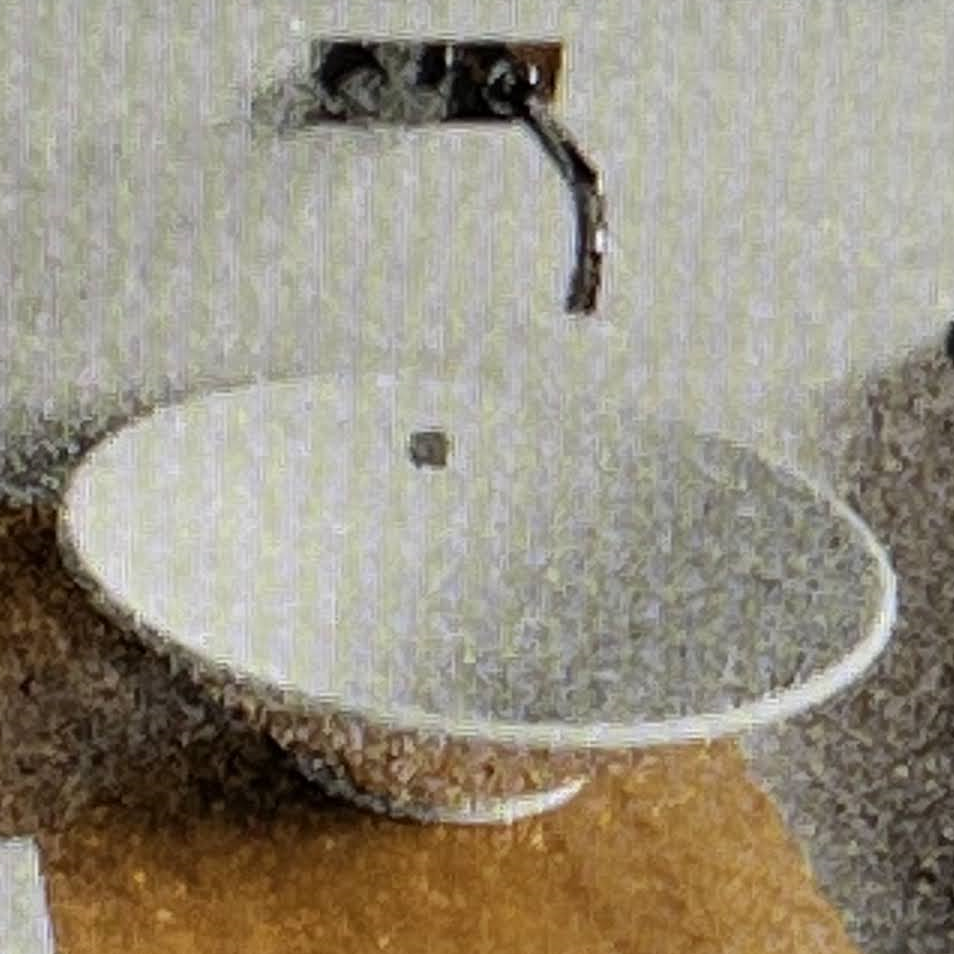}
    \includegraphics[width=0.31\columnwidth]{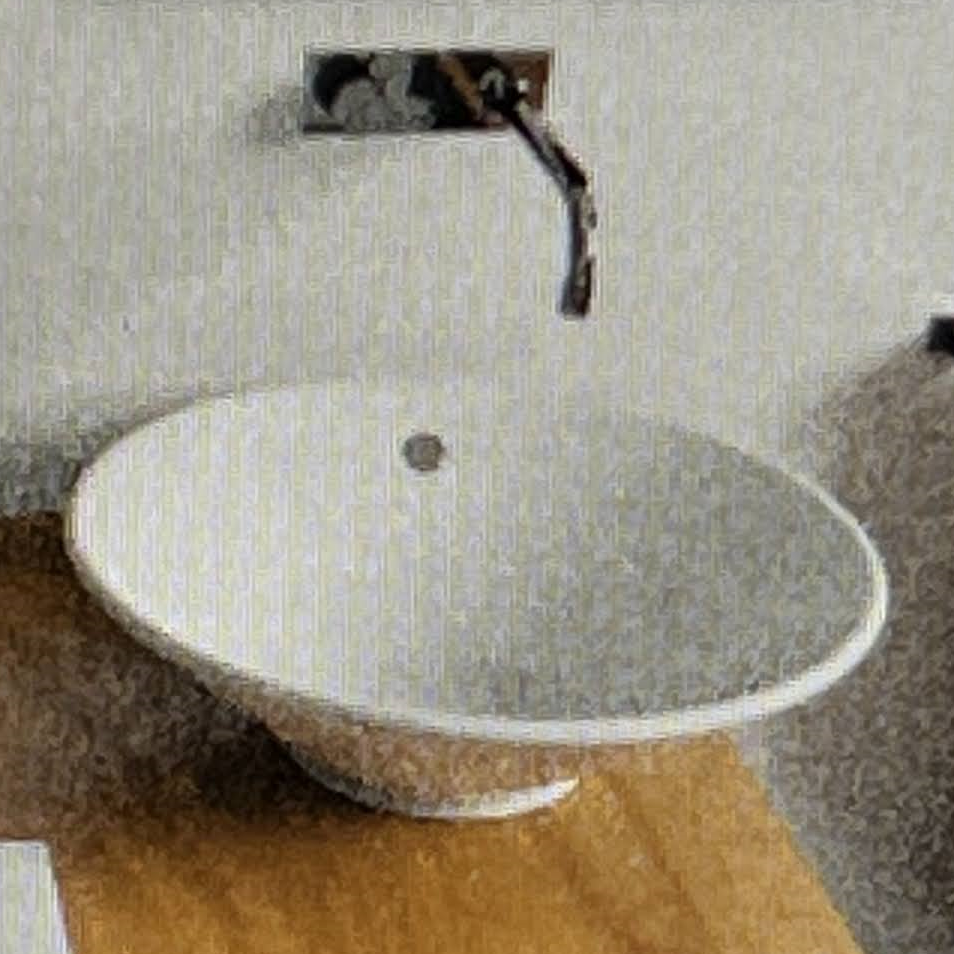}
    \caption[Caption]{Example of Monte Carlo noise (left) being reduced at 90 Hz (middle) and 360 Hz (right). Photos taken by iPhone from an animated demo\footnotemark}
    \label{fig:240hz}
\end{figure}

Enderton and Wexler established a useful "workflow scale" stating that as tasks vary in their completion time, the style of user input and workflow shifts~\cite{enderton2011workflow}. 
Tasks move from background and nightly batches that may take hours, days, or weeks; to direct, fluid and interactive tasks that operate at 10 to 100 Hertz (Hz). 
Esports and other applications requiring very rapid updates are driving the development of devices supporting this emerging interaction style. Until recently, computer monitors only operated at 60 Hz, but the demands of competitive gamers have pushed the market to release 360 and even 500 Hz monitors.
Recent research~\cite{andersson2019temporally} demonstrates that high-frame-rate interaction styles open opportunities for novel rendering approaches.
For example, in Figure~\ref{fig:240hz} we can see how noisy frames can be perceived as much lower noise levels when the frame rate is high enough.
% A similar example in Figure~\ref{fig:240hz2} shows how an aliased line similarly appears anti-aliased despite no such anti-aliasing occurring in the source image. 

While relatively little work has been done in graphics research at high refresh rates to this point, the future of real-time rendering may well head toward less strictly framed \cite{woolley2003interruptible} or frameless approaches \cite{bishop1994frameless}, where individual light samples are presented to the user as soon as they are available in the system.
This continuous update would more closely match the real-world physics of light, and could reduce rendering latency further.
The opportunities for novel rendering research in this space are enormous, and barely any research on the topic has been published yet.

It is worth noting that there is fairly clear evidence that even relatively small increases in refresh rate easily reach the noticeable threshold~\cite{denes2020perceptual,deber2015much}.
However, in studies seeking a player performance benefit, frame rate has shown marginal benefits for a small subset of aiming tasks thus far~\cite{spjut2019latency}.
More study is needed given the recency of variable refresh rate technology and the difficulty of studying refresh rate in isolation from latency given how closely they are coupled functionally in game implementations.

\footnotetext{ \url{https://developer.download.nvidia.com/research/temporally-dense-ray-tracing/bathroom.html} }

\section{Low latency rendering}

Interactive applications can be miserable to use when latency is high.
Systems with high network latency are particularly undesirable without latency hiding.
The latency between game client and server is such a well-known problem that games implement a wide array of latency-hiding techniques~\cite{liu2022survey}.
These latency hiding and compensation algorithms have come to be known colloquially as \textit{netcode}.
While many of the techniques listed in Liu's survey paper require a full game client to accomplish, one lighter weight approach to hide latency is VR-style late input sampling and image-based warping~\cite{kim2020post,boudaoud2021gaming}.
It is worth noting that client-server latency hiding techniques are quite effective, leaving local system latency as a much larger penalty on game performance than the same amount of network latency~\cite{liu2021comparing}.

For local computer system latency the primary contributors are input devices, game simulation, game rendering, and display output.
Rendering is a significant portion of the computer system latency for many computer games and happens to be the place where graphics researchers can contribute with novel algorithms.
Of particular interest might be methods for overlapping non-critical computation such that scene updates that are not fundamental to responsiveness or handling input can be handled out of the critical path. 
In addition to novel algorithms research, there are also some simple rendering pipeline changes that will reduce input to output latency, such as NVIDIA Reflex, a frame timing assistance development kit~\cite{nvidia_2020}.
In the local computer latency case, small steps in already low latency have been shown to still impact player performance for FPS aiming~\cite{spjut2021case}.
In a larger study, local latency was shown to have a significant effect on aiming performance across all skill levels, with up to a 2x increase in score when playing at 25 ms average latency as compared to 85 ms~\cite{spjut2022improving}.

In order to optimize latency, it is essential to have tools for measuring it.
Profiling tools are useful for finding functions that consume large amounts of CPU cycles, but it can be more challenging to capture and analyze whether a particular function is in the critical path for latency.
A common approach to end-to-end latency monitoring is to use a high speed camera and a LED attached to a mouse or keyboard button, counting frames from the LED lighting to screen update, but this approach is particularly labor intensive when capturing large numbers of measurements.
The Reflex Latency Analyzer (RLA) present in many G-SYNC monitors can be used as a lower friction approach for measuring click to photon latency.
Additionally, a variety of open source or more available tools have been released to measure latency effectively~\cite{schmid2021yet,dossena2022openldat}.

% \cite{Spjut19FPSci}

\section{Player performance and visual perception}

Since player performance is an important metric for any of these approaches, researchers can use games directly to analyze metrics such as wins and losses, kill/death ratios, actions per minute (APM) or other reported metrics.
However, we have found purpose-built aim trainers, such as KovaaK's~\cite{kovaaks} and Aim Lab~\cite{aimlab}, to be useful since they isolate most of the gameplay to focus on particular low level tasks, such as aiming.
We built such a tool, FirstPersonScience~\cite{Spjut19FPSci,boudaoud2022firstpersonscience}, and released it on GitHub.
One recent study focused on saccadic latency as a proxy for human reaction time in a gaming context~\cite{duinkharjav2022image}.
Such low level metrics should be balanced with more game-focused measures to avoid optimizing for the wrong things.

% TODO: Joohwan?

When performing tasks on an interactive system, the user regularly goes through perception-action cycles. 
Competition draws players to esports games leading them to care about perceptual elements that are crucial to their in-game performance. 
As a result, esports players tune graphics settings in ways that are sometimes unintuitive in the traditional view.

When playing esports action games, many players prioritize lowering latency ahead of most other visual settings since it gives them the best chance of winning. 
It is important to them to accelerate the perception-action cycle as much as possible by receiving near immediate visual feedback. 
High spatial resolution comes next as it visualizes spatial alignments more precisely. 
As a result, anti-aliasing settings are generally desirable as long as they do not compromise the game latency. 
Other high fidelity effects such as realistic light transport, shadows, and sub-surface scattering come last. 
It is likely an oversimplification to say that esports players do not care about high fidelity graphics at all. 
Rather, it reflects the emphasis on performance in esports players' merit function: it hurts far more to lose than to see lower fidelity visuals.

\section{Conclusions}
We are excited about the prospect of new esports research directions for graphics researchers.
The aspects of low latency, high frame rate, and human perception for user performance are relatively unexplored, particularly in comparison to the massive push toward photo realism necessitated by film and image quality focused games.
This expanding research niche presents opportunities for researchers to provide new tooling and equipment for developers, gamers, teams, and spectators alike to better enjoy competitive gaming.

\begin{acks}
    The authors would like to thank Mark Kilgard for helpful revisions of the manuscript.
\end{acks}